\documentclass[12pt]{article}
\pdfoutput=1
\usepackage{amsmath,amsfonts,graphicx,color,bbm,tikz,float}
\usepackage{comment}
\usepackage[nosort]{cite}
\usepackage{subfigure}
\usetikzlibrary{calc,positioning}
\usetikzlibrary{patterns,arrows,decorations.pathreplacing}
\usepackage{caption}
\usepackage{ulem}
\tikzset{>=stealth}
\usepackage{lipsum}

\newcommand\blfootnote[1]{%
  \begingroup
  \renewcommand\thefootnote{}\footnote{#1}%
  \addtocounter{footnote}{-1}%
  \endgroup}

\makeatletter\@addtoreset{equation}{section}\makeatother
\newcommand{\be}{\begin{equation}}
\newcommand{\ee}{\end{equation}}
\def\beq{\begin{equation}}
\def\eeq{\end{equation}}
\newcommand{\bea}{\begin{eqnarray}}
\newcommand{\eea}{\end{eqnarray}}

\newcommand{\ket}[1]{{\left| {#1} \right>}}

\def\nn{\nonumber}

\renewcommand{\title}[1]{\vbox{\center\LARGE{#1}}\vspace{3mm}}
\renewcommand{\author}[1]{\vbox{\center#1}\vspace{3mm}}

\newcommand{\email}[1]{\vbox{\center\tt#1}\vspace{3mm}}

\begin{document}
\begin{titlepage}
\begin{center}

{\large {\bf Generating W states with braiding operators} }

\author{Pramod Padmanabhan,$^a$ Fumihiko Sugino,$^b$ Diego Trancanelli$^{c,\star}$}\blfootnote{${}^\star$ On leave of absence from the Institute of Physics at the University of S\~ao Paulo, S\~ao Paulo, Brazil.}

\vskip -1cm
{$^a${\it Center for Theoretical Physics of Complex Systems,\\ Institute for Basic Science, Daejeon, South Korea}}
\vskip0.1cm
{ $^b${\it Center for Theoretical Physics of the Universe,\\
Institute for Basic Science, Daejeon, South Korea} 
\vskip0.1cm
$^c${\it Dipartimento di Scienze Fisiche, Informatiche e Matematiche, \\
Universit\`a di Modena e Reggio Emilia, via Campi 213/A, 41125 Modena, Italy \\ \& \\
INFN Sezione di Bologna, via Irnerio 46, 40126 Bologna, Italy}}
\email{pramod23phys, fusugino, dtrancan@gmail.com}

\vskip 1cm 
\end{center}

\abstract{
\noindent 
Braiding operators can be used to create entangled states out of product states, thus establishing a correspondence between topological and quantum entanglement. This is well-known for maximally entangled Bell and GHZ states and their equivalent states under Stochastic Local Operations and Classical Communication, but so far a similar result for W states was missing. Here we use generators of extraspecial 2-groups to obtain the W state in a four-qubit space and partition algebras to generate the W state in a three-qubit space. We also present a unitary generalized Yang-Baxter operator that embeds the W$_n$ state in a $(2n-1)$-qubit space. 
}

\end{titlepage}

\section{Introduction}

The non-local nature of quantum entanglement suggests a correspondence between the topological entanglement encoded in knots, links and related objects and the quantum entanglement of quantum states, as proposed in early and recent works including \cite{pk, lh1, as, ste , lh2, qa}. Efforts to validate this intuition have focused on the search for braiding operators to create entangled states out of product states \cite{lh3, lh4, lh5}. Braiding operators that are unitary enjoy special significance, as they also serve as quantum gates, but non-unitary operators have also been considered. Central to the task of finding braiding operators is the systematic construction of solutions to the $(d,m,l)$-{\it generalized Yang-Baxter Equation} (gYBE) \cite{er1, er2} 
\begin{equation}\label{gybe}
\left(R\otimes I^{\otimes l}\right)\left(I^{\otimes l}\otimes R\right)\left(R\otimes I^{\otimes l}\right) = \left(I^{\otimes l}\otimes R\right)\left(R\otimes I^{\otimes l}\right)\left(I^{\otimes l}\otimes R\right),
\end{equation}
where the invertible operator $R : V^{\otimes m} \rightarrow V^{\otimes m}$ acts on $m$ copies of a local Hilbert space $V$ of dimension $d$. Equation (\ref{gybe}) reduces to the usual Yang-Baxter Equation (YBE) when $m=2$ and $l=1$. Invertible solutions $R$ to (\ref{gybe}) are called {\it generalized Yang-Baxter operators} (gYBOs)~\cite{er1,er2}.  

The connection with topology arises from the fact that representations $\rho(\sigma_i)$ of the generators $\sigma_i$ of the {\it Artin Braid Group} can be built using the gYBOs, as follows: 
\begin{equation}
\rho(\sigma_i) = \left(I^{\otimes l}\right)^{\otimes i-1}\otimes R_{i,\cdots, i+m-1}\otimes \left(I^{\otimes l}\right)^{\otimes n-i-m+1},
\end{equation}
with the $\sigma_i$ satisfying the {\it braid relation}, $\sigma_i\sigma_{i+1}\sigma_i = \sigma_{i+1}\sigma_i\sigma_{i+1}$. 
One has also to guarantee that the {\it far-commutativity} condition $\sigma_i\sigma_j= \sigma_j\sigma_i$ for $|i-j|>1$  is satisfied, which happens automatically for $\frac{m}{2}\leq  l< m$ \cite{pfd1}.\footnote{Note that there exists no nontrivial gYBO for $l\geq m$. In that case, (\ref{gybe}) is solved only when $R$ is a projection operator. However, that is not invertible except for the trivial case 
$R={\bf 1}$. When (\ref{gybe}) depends on a spectral parameter we can find non-trivial invertible $R$ matrices for $l\geq m$ \cite{pfd2}.}  Here we shall focus our attention on this particular case. 

Several solutions to the gYBE in (\ref{gybe}) have been found in \cite{wk, rc, cg} and more recently using partition algebras in \cite{pfd1}. 
Among the quantum states obtained from these operators, one finds maximally entangled states such as the two-qubit Bell states, of typical form $\frac{1}{\sqrt{2}}[\ket{00}+\ket{11}]$, and the three-qubit GHZ states, of typical form $\frac{1}{\sqrt{2}}[\ket{000}+\ket{111}]$. 
In the classification by equivalence under {\it Stochastic Local Operations and Classical Communication} (SLOCC)\footnote{Two quantum states in an $n$-party system, 
$\ket{\psi_1}$ and $\ket{\psi_2}$ are SLOCC equivalent if they related by an {\it invertible local operator} (ILO), 
$\ket{\psi_1} = \left(A_1\otimes\cdots\otimes A_n\right)\ket{\psi_2}$. Here $A_i$ are invertible operators acting on the $i$th Hilbert space, $\mathcal{H}_i$ in the $n$-party Hilbert space.}~\cite{dur}, 
other classes of states such as partially entangled states and product states can also be obtained from these braiding operators. However, braiding operators creating W states like $\frac{1}{\sqrt{3}}[\ket{001}+\ket{010}+\ket{100}]$ remained elusive. 

Notice that the GHZ and W states are genuinely tripartite entangled states belonging to different SLOCC equivalence classes, because there is no ILO which transforms one into the other.  
Properties of these states are quite different. The GHZ states become unentangled bipartite mixed states after tracing out one of the qubits. On the other hand,  the  bipartite mixed states of W states are entangled (see \cite{walter_etal}, for example). This can be also seen through tangle invariants: the GHZ states give the maximal 3-tangle with vanishing 2-tangles, whereas the 3-tangle vanishes for W states~\cite{carteret,sudbery}. In this sense, W states are robust against particle loss and important for example in quantum memories, while GHZ states are fragile and with possible application to quantum secret sharing. 

In these notes, we obtain W states from the braiding operators constructed in \cite{pfd1} using partition algebras, something that was left out of that reference, and also present a novel construction in terms of extraspecial 2-groups \cite{er1} adapted to a four-qubit setting. It turns out that these methods only provide non-unitary braiding operators. We also embed the W state in a five-qubit space and its generalization -- the W$_n$ state typically given by 
$\frac{1}{\sqrt{n}}\left[\ket{100\cdots 0}+\ket{010\cdots 0}+\cdots +\ket{000\cdots 1}\right]$ -- in a $(2n-1)$-qubit space 
with unitary $(2,5,1)$- and $(2,2n-1,1)$-gYBOs, respectively. However, these fail to satisfy far-commutativity and hence are not braiding operators.


\section{Braiding operators from extraspecial 2-groups}\label{e2g}

The generators $\theta_j$ of extraspecial 2-groups satisfy  
\begin{equation}\label{e2grelations}
\theta_j^2 = -1,~~\theta_j\theta_{j+1}=-\theta_{j+1}\theta_j,~~\theta_j\theta_k=\theta_k\theta_j,~|j-k|>1.
\end{equation}
A representation of these generators which is adapted to studying a two-qubit space is given by 
\begin{equation}
\theta_j = \mathrm{i}X_jZ_{j+1} 
\end{equation}
with $X_j$ and $Z_{j+1}$ denoting the Pauli matrices acting on the $j$-th and $(j+1)$-th qubits respectively. This is  useful for finding W states, by considering 
the following ansatz for a gYBO on the four qubits $j,\cdots, j+3$:
\begin{equation}\label{R4ansatz}
R_j = 1 + \alpha_1 \theta_j + \alpha_2 \theta_{j+1} + \alpha_3 \theta_{j+2} + \beta_1 \theta_j\theta_{j+1} + \beta_2 \theta_{j+1}\theta_{j+2} + \beta_3 \theta_j\theta_{j+2} + \gamma \theta_j\theta_{j+1}\theta_{j+2},
\end{equation}
with complex parameters $\alpha_1, \alpha_2, \alpha_3, \beta_1, \beta_2, \beta_3$ and $\gamma$. 
Note that (\ref{R4ansatz}) can entangle the first three of the four qubits. 
The last qubit at $j+3$ is essentially left unchanged, since $R_j$ acts on it as a diagonal matrix.  
For example,
\bea
R_j\ket{0000} & = & \left[\ket{000} + \mathrm{i}\alpha_1\ket{100}+ \mathrm{i}\alpha_2\ket{010}+ \mathrm{i}\alpha_3\ket{001} \right. \nn \\
& & \left. +\beta_1\ket{110}+ \beta_2\ket{011}-\beta_3\ket{101} -\mathrm{i}\gamma\ket{111}\right]\otimes\ket{0}.
\eea

For simplicity, we limit ourselves to subsets of all the parameters appearing in the $(2,4,2)$-gYBE. 
Each term in (\ref{R4ansatz}) produces one of the eight product basis states of a three-qubit space, so that choosing subsets amounts to choosing gYBOs that create SLOCC-equivalent states to the standard $\frac{1}{\sqrt{3}}\left[\ket{001}+\ket{010}+\ket{100}\right]$. There are five possibilities where the W state SLOCC class comprises either four, five, six, seven or eight of the product basis states. Out of these we found that there is no non-trivial gYBO that produces entangled states made of four product basis states that is SLOCC equivalent to the W state, so we leave out this possibility. The case when the entangled states is a superposition of all the eight product basis states is computationally hard to solve and we leave this out as well. We find non-unitary gYBOs that produce entangled states consisting of superpositions of five, six and seven product basis states.

\subsection{Superposition of 5 states}

There are five inequivalent (non-unitary) solutions, that we consider separately. 

\paragraph{Case 1: $\beta_2=-\beta_1, ~\beta_3=-\beta_1^2,~\gamma=\mathrm{i}\sqrt{\left(1+\beta_1^2\right)^2},~\alpha_1=\alpha_2=\alpha_3=0$.}
There are two inequivalent non-unitary solutions in this case. For $\beta_1=\frac{\mathrm{i}}{\sqrt{3}}$ and $\beta_1=1$ we obtain $(2,4,2)$-gYBOs that produce the W-state equivalents of five states in the superposition. 
Their eigenvalues are in the sets $\{1_{(8)}, \pm \frac{1}{3}_{(4)}\}$ and $\{-1_{(4)}, 1_{(12)}\}$, respectively. The notation $e_{(k)}$ denotes an eigenvalue $e$ with multiplicity $k$. The eigenvalues of the former set do not have the same modulus, whereas those of the latter are phases and could possibly be mapped to a unitary braiding operator via an ILO.\footnote{
It is not possible to make the former set unitary via an ILO, since ILOs do not change the eigenvalues.}
However this is not what happens, as the eigenvectors of the second braiding operator do not span the complete basis. The operator is not diagonalizable and can take, at most, Jordan canonical form. 

\paragraph{Case 2: $\alpha_2=\mathrm{i}\sqrt{\left(1+\beta_1^2\right)^2}, ~\beta_2=-\beta_1,~ \beta_3=-\beta_1^2,~ \alpha_1=\alpha_3=\gamma=0$.}
We obtain a $(2,4,2)$-gYBO that yields a W-state equivalent for $\beta_1=\mathrm{i}\sqrt{3}$ with eigenvalues $\{1_{(8)}, \pm 3_{(4)}\}$. Clearly this is a non-unitary braiding operator that cannot be mapped to a unitary braiding operator with an ILO. 

\paragraph{Case 3: $\alpha_1=\alpha_3=\sqrt{\alpha_2\left(\alpha_2-\mathrm{i}\right)},~ \gamma=-\alpha_2+\mathrm{i},~ \beta_1=\beta_2=\beta_3=0$.}
For $\alpha_2=\frac{1}{2}$ this gives a non-unitary braiding operator constructed from a $(2,4,2)$-gYBO with eigenvalues $\{\left(1+\frac{\mathrm{i}}{2}\right)_{(8)}, -\frac{\mathrm{i}}{2}_{(8)}\}$ 
that generates a W-state equivalent.

\paragraph{Case 4: $\alpha_1=\sqrt{-\frac{1}{2}+\alpha_2^2-\frac{1}{2}\sqrt{1-4\alpha_2^2}},~ \alpha_3=\frac{\alpha_2^2}{\alpha_1},~ \gamma=-\alpha_2,~ \beta_1=\beta_2=\beta_3=0$.}
This gives a non-unitary braiding operator generating a state in the W-state class for $\alpha_2=\frac{1}{2}$, with eigenvalues $\{1_{(16)}\}$. Though the eigenvalues are phases, this operator cannot be mapped to a unitary operator by an ILO as 
the eigenvectors do not span the complete basis and hence cannot be diagonalized.

\subsection{Superposition of 6 states}\label{s6}

There are three possible non-unitary solutions which we consider separately.

\paragraph{Case 1: $\alpha_2=\alpha_1, ~\beta_1=\beta_2=-\mathrm{i},~\beta_3=-1,~\alpha_3=\gamma=0$.}

This one-parameter $(2,4,2)$-gYBO yields an entangled state that is SLOCC-equivalent to the W state for $\alpha_1=-\mathrm{i}\sqrt{2}$. 
Up to normalization, the eigenvalues of this operator are $\{ \pm 1_{(4)}, \left(1\pm\sqrt{2}\right)_{(4)}\}$.  As these absolute values are not the same, we do not expect to find an ILO that maps this to a unitary operator. 

\paragraph{Case 2: $\alpha_1=\alpha_3=-\mathrm{i}\beta_1, ~\beta_2=-\beta_1,~\gamma=\mathrm{i}\sqrt{1+4\beta_1^2}, \alpha_2=\beta_3=0$.}

This time we obtain a $(2,4,2)$-gYBO that yields a W-state equivalent for $\beta_1=-\frac{\mathrm{i}}{\sqrt{2}}$ with eigenvalues $\{e^{\pm\mathrm{i}\frac{\pi}{4}}_{(8)}\}$. 
As they are phases, one could expect to find an ILO that maps this to a unitary braiding operator. However it turns out not to be the case, as we show now.
Consider the matrix $V$ formed by the eigenvectors of the gYBO in question. The non-normalized eigenvectors can be split into two sets given by
\begin{eqnarray}
 & & \ket{v_1} = \ket{0001} + \ket{1111}, \quad \ket{v_2} = \ket{1110} - \ket{0000}  , \quad
\ket{v_3} = \ket{1101} - \ket{0011}  , \nn \\
& & \ket{v_4} = \ket{0010} + \ket{1100}, \quad 
\ket{v_5} = \ket{1011} - \ket{0101}  ,\quad \ket{v_6} = \ket{0100} + \ket{1010}, \nonumber \\
& & \ket{v_7} = \ket{0111} + \ket{1001}, \quad \ket{v_8} = \ket{1000} - \ket{0110}  , 
\end{eqnarray}
and 
\begin{eqnarray}
\ket{v_9} & = & \sqrt{2} (\ket{0011} - \ket{0111}) - \ket{0001} + \ket{1111}, ~~ \ket{v_{10}}  =   \ket{0000} + \sqrt{2} (\ket{0010} -\ket{0110}) + \ket{1110},  \nonumber \\
 \ket{v_{11}} & = &  \ket{0011} - \sqrt{2} (\ket{0001} +\ket{0101}) + \ket{1101}, ~~ \ket{v_{12}}  = \ket{1100} - \sqrt{2} (\ket{0100} +\ket{0000}) - \ket{0010}, \nonumber \\
\ket{v_{13}} & = & \sqrt{2} (\ket{0011} - \ket{0111}) + \ket{0101} + \ket{1011}, ~~ \ket{v_{14}}  =  \sqrt{2} (\ket{0010} - \ket{0110}) - \ket{0100} + \ket{1010}, \nonumber \\
 \ket{v_{15}} & = & \sqrt{2}( \ket{0001} +  \ket{0101}) - \ket{0111} + \ket{1001}, ~~ \ket{v_{16}}  =  \sqrt{2} (\ket{0000} +  \ket{0100}) + \ket{0110} + \ket{1000}. \nonumber \\
\end{eqnarray}
Then
\be
V=\left(\ket{v_1},\cdots, \ket{v_8},\ket{v_9},\cdots, \ket{v_{16}}\right),
\label{V}
\ee
with each of $\ket{v_i}$ regarded as a column vector, has full rank and is invertible. However, as can be seen by inspection, these eigenvectors do not form an orthonormal set and hence $V$ is not unitary. 
The gYBO is expressed as $R_j=V D V^{-1}$, with $D$ a diagonal matrix given by the eigenvalues.   
Consider the polar decomposition of the matrix $V=HU$, with $H$ a positive definite hermitian matrix and $U$ a unitary matrix. We can ask the question whether $QV$, with $Q$ an ILO, results in a unitary matrix. If this were true, then $Q$ should be of the form $Q=WH^{-1}$, with $W$ another unitary matrix, and $Q^\dag Q= H^{-2}=\left(VV^\dag\right)^{-1}$. 
%
Being an ILO, $Q$ can be written as $Q_1\otimes Q_2\otimes Q_3\otimes Q_4$, with $Q_j$ an ILO acting only on the $j$-th qubit. 
We find that the tensor-product structure 
$Q^\dagger Q=\left(Q_1^\dag Q_1\right) \otimes \left(Q_2^\dag Q_2\right) \otimes \left(Q_3^\dag Q_3\right) \otimes \left(Q_4^\dag Q_4\right)$ 
is not compatible with $\left(VV^\dag\right)^{-1}$ computed from (\ref{V}). This shows that there exists no ILO $Q$ that maps the non-unitary $(2,4,2)$-gYBO in question into a unitary operator.   


\paragraph{Case 3: $\alpha_2=-\mathrm{i}, ~\beta_2=-\beta_1,~\beta_3=1-\frac{\beta_1^2}{2},~~\gamma=-\frac{\mathrm{i}}{2}\left(2+\beta_1^2\right), \alpha_1=\alpha_3=0$.}
We now have a $(2,4,2)$-gYBO which for $\beta_1=-2k$ yields an entangled state in the W-state SLOCC class. This operator has eigenvalues $\{1_{(8)},\pm\left(1-2k^2\right)_{(4)}\}$, with $k$ a non-zero parameter. In general these are non-unitary solutions as the eigenvalues are not phases except for $k=\pm 1$. However, this operator has eigenvectors that contain components of the form, $\frac{2k}{k^2-1}$ and $\frac{k^2+1}{k^2-1}$ which become singular at $k=\pm 1$.
Although the normalized eigenvectors are not singular in the limit $k\to 1$ or $-1$, they do not span the complete basis and the operator is not diagonalizable. 
For all other non-zero values we obtain non-unitary braiding operators. 

\subsection{Superposition of 7 states}

We obtain a non-unitary braiding operator built from a gYBO satisfying the $(2,4,2)$-gYBE when the parameters satisfy
\begin{equation}\nonumber
\alpha_1=\alpha_3=\sqrt{\alpha_2^2-\beta^2_2-\mathrm{i}\alpha_2\sqrt{1+4\beta_2^2}}, ~~ \beta_1=-\beta_2,~~\gamma=-\alpha_2+\mathrm{i}\sqrt{1+4\beta_2^2},~~\beta_3=0,
\end{equation}
for $\alpha_2=1,~ \beta_2 = -2^{-\frac14}\cdot e^{-\mathrm{i}\frac{\pi}{4}}$. 
$\{\left(1+\sqrt{2}+\mathrm{i}\right)_{(8)}, \left(1-\sqrt{2}-\mathrm{i}\right)_{(8)}\}$ are the eigenvalues of this operator, revealing its non-unitary character. 

\section{Braiding operators from partition algebras}\label{part}

Braiding operators were systematically constructed in \cite{pfd1} out of generators of partition algebras. We do not go over the construction here, but refer the reader to \cite{pfd1, ram} for details. We use two types of 3-qubit braiding operators from \cite{pfd1} for the purpose of obtaining W states. 

\subsection{Using the generators $s_j$, $p_j$, $p_{j+1}$, $p_{j+2}$}

We first use $s_j$, the permutation operator that swaps the indices $j$ and $j+1$ and the projector $p_j$, for which we use the representation $p_j = \frac{1+X_j}{2}$, with $X$ being the first Pauli matrix. 
Then the operators  
\begin{equation}\label{nb31}
R_j = s_{j, j+2}\left(1 + \alpha_1~p_j + \alpha_3~p_{j+2} + \beta_1~p_jp_{j+1} + \beta_2~p_{j+1}p_{j+2} +  \beta_3~p_jp_{j+2}
+ \gamma~p_jp_{j+1}p_{j+2}\right),
\end{equation}
with $s_{j,j+2}=s_js_{j+1}s_j$, satisfy the $(2,3,2)$-gYBE for 
\begin{equation}\label{nsol1}
\beta_2 = -\frac{\beta_1\left(1+\alpha_3\right)}{1+\alpha_1+\beta_1},~~\gamma=\frac{\beta_1\left(\alpha_3-\alpha_1-\beta_1\right)}{1+\alpha_1+\beta_1}.
\end{equation}
The SLOCC classes of the states obtained from the unitary points of these operators are discussed in \cite{pfd1}. Here we look at points in the parameter space that yield W states.

In terms of parameters $\beta_1$, $l_1$ and $l_3$ we find that at
\begin{equation}
\alpha_1=-\frac{\beta_1}{2}-\frac{2}{l_1+1},~\alpha_3=\frac{\left(l_1+1\right)\left(l_3-1\right)\beta_1-4\left(l_1-1\right)}{2\left(l_1-1\right)\left(l_3+1\right)},~\beta_3=\frac{\left(l_1-l_3\right)\left(l_1+1\right)\beta_1+4\left(l_1-1\right)}{\left(l_1+1\right)\left(l_1-1\right)\left(l_3+1\right)},
\end{equation}
the gYBO in (\ref{nb31}) maps the product state $\left(l_1\ket{0}+\ket{1}\right)\otimes\ket{1}\otimes\left(l_3\ket{0}+\ket{1}\right)$ to the state $k_1\ket{001}+k_2\ket{010}+k_3\ket{100}$ with $k_1=\frac{1}{4}\left(l_1+1\right)\left(l_3-1\right),$ $k_2=\left(l_1-1\right)\left(l_3-1\right)$ and  $k_3=-\frac{1}{4}\left(l_1+1\right)\left(l_3-1\right)\beta_1$. 
The four eigenvalues of this operator are 
\begin{equation}\nonumber
\left\{1_{(2)}, \frac{\left(l_1-1\right)\left(l_3-1\right)}{\left(l_1+1\right)\left(l_3+1\right)}_{(2)}, \pm\mathrm{i}\frac{\sqrt{l_3^2-1}\sqrt{\left(l_1+1\right)^2\beta_1^2 - 4\left(l_1-1\right)^2}}{2\left(l_3+1\right)\sqrt{l_1^2-1}}_{(2)}\right\},
\end{equation}
which are in general not phases, thus making the braiding operator non-unitary. However at $l_3 = \frac{2\left(1-l_1\right)}{1-l_1+e^{\mathrm{i}\theta}\left(1+l_1\right)}-1$ and $\beta_1= \frac{2\left(l_1-1\right)}{l_1+1} e^{-\mathrm{i}\frac{\theta}{2}}\sqrt{e^{\mathrm{i}\theta}-e^{2\mathrm{i}\phi}}$ the eigenvalues reduce to $e^{\mathrm{i}\theta}$ and $e^{\mathrm{i}\phi}$, opening up the possibility for the existence of an ILO that maps this non-unitary braiding operator to a unitary one. However, by using the arguments of Case 2 in Sec. \ref{s6} we find that such an ILO does not exist.  

\subsection{Using the generators $s_j$, $p_{j, j+1}$, $p_{j+1, j+2}$}

In this case we use the operators $p_{j, j+1}=1+X_jX_{j+1}$ along with the permutation operators to obtain the braiding operators of the form
\begin{equation}\label{nb32}
R_j = s_{j, j+2}\left(1 + \alpha~p_{j,j+1} + \beta~p_{j+1,j+2} + \gamma~p_{j,j+1}p_{j+1,j+2} + \delta~p_{j,j+2}\right),
\end{equation}
satisfying the $(2,3,2)$-gYBE when $\gamma=-\frac{\alpha+\beta}{2}$.

This operator yields a W-state equivalent for $\delta=-\frac{1}{2}\left(2+\alpha+\beta\right)$. It is easy to check that at this point it takes the product state $\ket{000}$ to $\left(\frac{\alpha-\beta}{2}\right)\ket{011}-\left(1+\alpha+\beta\right)\ket{101}-\left(\frac{\alpha-\beta}{2}\right)\ket{110}$, which is SLOCC-equivalent to the standard W state $\frac{1}{\sqrt{3}}\left[\ket{001}+\ket{010}+\ket{100}\right]$. 
The eigenvalues are given by 
\begin{equation}\nonumber
\left\{-\left(1+\alpha+\beta\right)_{(4)}, \pm\sqrt{\left(1+2\alpha\right)\left(1+2\beta\right)}_{(2)}\right\},
\end{equation}
which are not phases in general, except for $\alpha=\frac{1}{2}\left[-\left(e^{\mathrm{i}\phi}+1\right)\pm\sqrt{e^{2\mathrm{i}\phi}-e^{2\mathrm{i}\theta}}\right]$ and $ \beta=\frac{1}{2}\left[-\left(e^{\mathrm{i}\phi}+1\right)\mp\sqrt{e^{2\mathrm{i}\phi}-e^{2\mathrm{i}\theta}}\right]$. Again, the same arguments as in Case 2 of Sec. \ref{s6} rule out the possibility of an ILO mapping this to a unitary operator. 

\section{Unitary gYBOs as W-state entanglers} \label{URW}

So far, all the examples we have found were of non-unitary operators. Unitary solutions to gYBEs that generate W states from product states can be found, as we discuss now. First, let us introduce generators $\xi_j$ satisfying relations:
\bea
 & & \xi_j^2=-1,\qquad \xi_j\xi_{k}=-\xi_{k}\xi_j \quad (|j-k|=1,2), \nn \\
& & \xi_j\xi_{k}=\xi_{k}\xi_j \quad (|j-k|>2),
\label{e2g_like}
\eea
which are similar to the relations of the extraspecial 2-group (\ref{e2grelations}). We pick a realization on a three-qubit space for each $\xi_j$: 
\be
\xi_j=\mathrm{i}X_jZ_{j+1}Z_{j+2},
\label{xi}
\ee
and consider the operator acting on five qubits at $j,\cdots, j+4$: 
\be
R_j=1+\alpha\xi_j+\beta\xi_{j+1}+\gamma\xi_{j+2},
\label{unitaryR}
\ee
where $\alpha$, $\beta$ and $\gamma$ are parameters. Since $\xi_j$, $\xi_{j+1}$ and $\xi_{j+2}$ are antihermitian and anticommute with each other, 
the parameters should be real in order for (\ref{unitaryR}) to be unitary (up to overall normalization). 
It is easy to see that 
\be
R_j\ket{00000}=\left[\ket{000}+\mathrm{i}\alpha\ket{100}+\mathrm{i}\beta\ket{010}+\mathrm{i}\gamma\ket{001}\right]\otimes\ket{00}.
\ee
For $\alpha$, $\beta$ and $\gamma$ nonzero, the first 3-qubits on the RHS are SLOCC-equivalent to the standard W state.\footnote{For example, the equivalence can be seen via the ILO \cite{dur}
\[
\begin{pmatrix}1 & \mathrm{i}/\alpha \\ 0 & -\mathrm{i}/\alpha \end{pmatrix}\otimes  \begin{pmatrix}1 & 0 \\ 0 & -\mathrm{i}/\beta \end{pmatrix}\otimes 
\begin{pmatrix}1 & 0 \\ 0 & -\mathrm{i}/\gamma \end{pmatrix}_.
\]
} 
We recognize the last two qubits as a spectator state.

By using (\ref{e2g_like}), it is straightforward to find that the $R$-matrix (\ref{unitaryR}) with all the parameters different from zero satisfies the $(2,5,1)$-gYBE 
if and only if $\alpha=\beta=\gamma=\pm1/\sqrt{5}$. 
(For $(2,5,l)$-gYBE with $l=2,3,4$, the ansatz (\ref{unitaryR}) does not provide a solution.) 
Including the overall normalization, we conclude that the $R$-matrices 
\be
\tilde{R}_j^{(\pm)}=\frac{\sqrt{5}}{2\sqrt{2}}\cdot{\bf 1}\pm \frac{1}{2\sqrt{2}}\left(\xi_j+\xi_{j+1}+\xi_{j+2}\right)
\label{unitaryR2}
\ee
generating the W state are unitary ($\tilde{R}_j\tilde{R}_j^\dagger={\bf 1}$) and satisfy the $(2,5,1)$-gYBE. They have the common  eigenvalues 
$\left\{\left(\frac{\sqrt{5}\pm \mathrm{i}\sqrt{3}}{2\sqrt{2}}\right)_{(16)}\right\}$. However, since (\ref{unitaryR2}) does not satisfy far-commutativity, namely $R_jR_{j+k} \neq R_{j+k}R_j$ for $k=2, 3, 4$, they do not form unitary representations of the braid group. 
Instead, $R_jR_{j+k}=R_{j+k}R_j$ for $k>4$ is satisfied. 
Nevertheless we observe that $\tilde{R}_j^{(\pm)}$ in (\ref{unitaryR2}) generates an infinite group as can be seen by computing the powers of $\tilde{R}_j^{(\pm)}$,
\be
\left(\tilde{R}_j^{(\pm)}\right)^n = \left(\frac{\sqrt{5}}{2\sqrt{2}}\right)^n\left\{\sum_{k=0}^{\lfloor \frac{n}{2}\rfloor} \begin{pmatrix} n \\ 2k\end{pmatrix} \left(-\frac35\right)^k\cdot {\bf 1} 
\pm\frac{1}{\sqrt{5}}\sum_{k=0}^{\lfloor\frac{n-1}{2}\rfloor}\begin{pmatrix} n \\ 2k+1\end{pmatrix} \left(-\frac35\right)^k \left(\xi_j+\xi_{j+1}+\xi_{j+2}\right)\right\}
\ee
and a similar expression for $\tilde{R}_j^{-1}$. $\lfloor x\rfloor$ represents the greatest integer not exceeding $x$. 
Here we have used the identity $\left(\xi_j+\xi_{j+1}+\xi_{j+2}\right)^2=-3$ which follows from the relations satisfied by $\xi$ in (\ref{e2g_like}). Clearly we generate an infinite group and thus quantum gates simulated using these operators can provide a universal set of gates for topological quantum computation \cite{er3}. 

It is possible to generalize this construction for the W$_n$ state, namely the W state for $n$ qubits. As an analog of extraspecial 2-group generators, one introduces 
\bea
 & & \eta_j^2=-1,\qquad \eta_j\eta_{k}=-\eta_{k}\eta_j \quad (|j-k|=1,\cdots, n-1), \nn \\
& & \eta_j\eta_{k}=\eta_{k}\eta_j \quad (|j-k|>n-1),
\label{e2g_like2}
\eea
and 
\be
R_j=1+\alpha\left(\eta_j+\cdots +\eta_{j+n-1}\right).
\label{unitaryRg}
\ee
For a realization of $\eta_j$ on $n$ qubits:
\be
\eta_j=\mathrm{i}X_jZ_{j+1}\cdots Z_{j+n-1},
\label{eta}
\ee
(\ref{unitaryRg}) is an operator on $(2n-1)$ qubits. When $\alpha$ is nonzero and real, $R_j$ is unitary and $R_j\ket{0}^{\otimes (2n-1)}$ is SLOCC-equivalent to the W$_n$ state with a spectator state 
of the last $(n-1)$ qubits $\ket{0}^{\otimes (n-1)}$. For $n=3$, this reduces to the case above. 

Equation (\ref{unitaryRg}) is shown to be a solution to the $(2,2n-1,1)$-gYBE for $\alpha=\pm 1/\sqrt{3n-4}$. With the overall normalization, the unitary $R$-matrices 
\be
\tilde{R}_j^{(\pm)} =\frac{\sqrt{3n-4}}{2\sqrt{n-1}}\cdot{\bf 1}\pm \frac{1}{2\sqrt{n-1}}\left(\eta_j+\cdots +\eta_{j+n-1}\right)
\label{unitaryRg2}
\ee
generating the W$_n$ state solve the $(2,2n-1,1)$-gYBE.  These cannot be regarded as braiding operators since the far commutativity relation does not hold 
(but $R_jR_{j+k}=R_{j+k}R_j$ ($k>2n-2$) is satisfied). 

\section{Conclusion}
\label{conclusion}

In these notes, we constructed gYBOs generating W states based on extraspecial 2-groups and partition algebras. The gYBOs are either non-unitary operators satisfying the far-commutativity relation or unitary operators not satisfying  far-commutativity. The former can be regarded as non-unitary braiding operators, namely non-unitary representations of the braid group. Though the latter do not generate the braid group they could satisfy braid-like relations and give rise to an infinite group related to the topology of the underlying space. We leave this for future investigation. 

One could apply the enhancing method described in \cite{turaev,hong} to the braiding operators obtained here and in \cite{pfd1}, thus producing link invariants corresponding to GHZ and W states, as well as partially entangled states. 
  
It seems quite hard to obtain unitary braiding operators for W states and we are not aware of any example in the literature. This is in sharp contrast to the case of GHZ states, for which various unitary representations have been found in \cite{er1, er2, pfd1}. It would be interesting to understand this difference better and to see whether it has any relevance in distinguishing between topological aspects of gYBOs and quantum entanglement.

\subsection*{Acknowledgements}
 PP and FS are supported by the Institute for Basic Science in Korea (IBS-R024-Y1, IBS-R018-D1). DT is supported in part by the INFN grant {\it Gauge and String Theory (GAST)}.


\end{document}